\documentclass[traditabstract]{aa}
\usepackage{graphicx}
\usepackage{txfonts}
\usepackage{natbib}
\def\crm{\cr\noalign{\medskip}}
\def\aa{-3}
\def\ab{-}
\def\ac{+9}
\def\ar{2}
\def\sign{\mathrm{sign}}
\def\dw{\delta}
\def\bfi{}
\begin{document}
  \title{On the equilibrium rotation of Earth-like extra-solar planets}

  \author{Alexandre C.M. Correia \inst{1,2}, Benjamin Levrard\inst{2,3},
         \and Jacques Laskar\inst{2}}

  \authorrunning{A.C.M. Correia, B. Levrard, and J. Laskar}

  \offprints{A.C.M. Correia, \email{correia@ua.pt}}

 \institute{Departamento de F\'isica da Universidade de Aveiro,
             Campus Universit\'ario de Santiago, 3810-193 Aveiro, Portugal
\and
           Astronomie et Syst\`emes Dynamiques, IMCCE-CNRS UMR 8028, 
             77 Avenue Denfert-Rochereau, 75014 Paris, France
\and
             Universit\'e de Lyon 1, CRAL,
  Ecole Normale Sup\'erieure de Lyon, 46 all\'ee d'Italie, 69364 Lyon Cedex 07, France}

  \date{Received ?? May, 2008; accepted ?? }

 \abstract{The equilibrium rotation of tidally evolved ``Earth-like''
 extra-solar planets is often assumed to be synchronous with their orbital mean
 motion. The same assumption persisted for Mercury and Venus until radar
 observations revealed their true spin rates.
 As many of these planets follow eccentric orbits and are believed to host
 dense atmospheres, we expect the equilibrium rotation to differ from the
 synchronous motion. Here we provide a general description of the allowed final
 equilibrium rotation states of these planets, and apply this to already discovered
 cases in which the mass is lower than 12 $M_{\oplus}$. At low
 obliquity and moderate eccentricity, it is shown that there are at most four distinct
 equilibrium possibilities, one of which can be retrograde. Because most
 presently known ``Earth-like'' planets present eccentric orbits, their
 equilibrium rotation is unlikely to be synchronous.} 


 \maketitle


\section{Introduction}

After a significant number of discoveries of new extra-solar gaseous giant planets, a new
barrier has been passed with the detections of several planets in the
Neptune and even Earth-mass ($M_{\oplus}$) regime: $5-12$ $M_{\oplus}$
\citep{Rivera_etal_2005,Lovis_etal_2006,Udry_etal_2007,Bonfils_etal_2007}.
If the commonly accepted core-accretion model can account
for the formation of these planets, resulting in a mainly icy/rocky composition, the
fraction of the residual He-H$_2$ atmospheric envelope accreted during the planet migration is
not tightly constrained for planets more massive than the Earth \citep[e.g.][]{Alibert_etal_2006}.
A minimum mass of below 10 $M_{\oplus}$ is usually considered to be the boundary between terrestrial
and giant planets, but \citet{Rafikov_2006} found that planets more massive than 6 $M_{\oplus}$ could have
retained more than 1 $M_{\oplus}$ of the He-H$_2$ gaseous envelope. For comparison, masses of Earth's
and Venus' atmosphere are respectively $\sim 10^{-6}$ and $10^{-4}$ times the planet's mass.
Despite significant uncertainties, these discoveries of ``super-Earths'' provide
an opportunity to test some properties that could be similar to those of our
more familiar telluric planets.

Because some of these planets are potentially in the ``habitable zone''
\citep{Udry_etal_2007,Selsis_etal_2007}, their spin state is an important
factor in understanding their climate. 
For planets with a negligible atmosphere such as Mercury, solid tides induced by the host star are expected
to despin the planet until an equilibrium position or a capture in a spin-orbit
resonance, depending on the orbital eccentricity and the permanent quadrupole
moment of inertia \citep{Goldreich_Peale_1966,Correia_Laskar_2004}. 
However, for planets with dense atmosphere such as Venus, thermal atmospheric
tides, driven by solar insolation, have a profound influence on the spin, and may destabilize
the previous tidal equilibrium state.
\citet{Correia_Laskar_2001,Correia_Laskar_2003I} investigated
the combined effect of gravitational and atmospheric tides on Venus
and showed that the planet could evolve into two different final states,
the existing retrograde motion being the most probable one.
This study was  performed with the small eccentricity approximation  
that may no longer be adequate for  ``Earth-like'' planets,
which exhibit a wide range of eccentricities, orbital distances,
or central star types. 
Following the work of \citet{Laskar_Correia_2004}, we generalize here 
these previous  studies and investigate the possible
equilibrium rotation states of ``Earth-like'' planets 
in the presence of significant eccentricity.
Although our knowledge of these planets is restricted to their orbital
parameters and minimum masses, we attempt to place new constraints on their
surface rotation rate, assuming that they have a dense atmosphere.

\section{Tidal evolution}

\label{tmodels}

Tidal effects are generated by differential and inelastic deformation of a planet
by a perturbing body.  For a planet with a dense atmosphere, we consider the
traditional gravitational tides ($ \tau=g $), and also thermal atmospheric tides
($ \tau=a $) produced by the stellar heating of the atmosphere.
In both cases, 
the averaged variation in the spin can be
expressed by the variables that characterize the spin (the rotation rate $ \omega $
and the obliquity $ \varepsilon $), and the elliptical elements, with
semi-major axis $ a $ and eccentricity $ e $
\citep[e.g.][]{Kaula_1964,Correia_Laskar_2003JGR}: 
\begin{equation}
\frac{d L}{d t} = T_\tau = K_\tau \sum_\sigma b_\tau (\sigma)
\Lambda_{\tau,\sigma} (e, \varepsilon) \ , \label{eq01}
\end{equation}
where $L=C \omega$ is the angular momentum of the planet, $C$ is the polar moment of inertia, 
the sum being taken over different harmonics of the tidal frequency $ \sigma$ (an integer combination of the
rotation rate $ \omega $ and the mean motion $ n $),
$ K_\tau $ is a factor related to the strength of the tide, $ b_\tau (\sigma) $
is an odd
function related to the dissipation within the planet, and $ \Lambda_{\tau,\sigma}
$ are polynomial functions of $ e $ and $ \cos \varepsilon $.
Since discovered Earth-like planets have moderate eccentricities (typically $e < 0.25$), we neglect terms
in $ e^4 $, which allows us to  simplify the tidal torques acting on the planet. In the same way,
we consider that the planet's obliquity $\varepsilon$ is small, neglecting terms higher than $ \varepsilon^2$ in
 Eq.(\ref{eq01}).


Gravitational tides are raised on the planet by the star because of the effect
of the gravitational gradient across the planet.
Equation (\ref{eq01}) becomes \citep[eg.][]{Kaula_1964}:
\begin{eqnarray}
T_g & = & \frac{3}{2} K_g \left[ \left(1-5 e^2\right) \, b_g (2 \omega - 2 n)
\phantom{\frac{1}{1}} \right. \crm
& & \left. + \frac{1}{4} \, e^2 b_g (2 \omega - n) +\frac{49}{4} \, e^2 b_g (2
\omega - 3 n) \right] \ , \label{eq02}
\end{eqnarray}
where $ K_g = - G M_*^2 R^5 / a^6 $, $ G $ is the gravitational constant, $ M_*
$ is the mass of the star, and $ R $ is the mean radius of the planet.
Since planets are not perfectly rigid, there
will be a distortion that produces a tidal bulge of amplitude $ k_2 $,
the second order potential Love number.
Imperfect elasticity delays the planet's response to the
perturbation by a time lag $ \Delta t_g $.
The deformation 
therefore lag behind the perturbation by 
an angle $ \delta_g = \sigma \Delta t_g /2 $ and
$ 
b_g (\sigma) = \! k_2 \! \sin 2 \delta_g  = k_2  \sin
\left( \sigma \Delta t_g  \right)  
$ 
\citep[e.g.][]{Correia_Laskar_2003JGR}.


The differential absorption of the stellar heat by the planetary atmosphere gives
rise to local variations in temperature and consequently to pressure gradients.
The mass of the atmosphere is then redistributed, adjusting for an
equilibrium position.
Observations on Earth show that the pressure redistribution is
essentially a superposition of two pressure waves: a diurnal tide of small
amplitude (with no dynamical counterpart)
and a strong semi-diurnal tide \citep[see][]{Chapman_Lindzen_1970}.
As for gravitational tides, the redistribution of mass in the atmosphere
produces an atmospheric bulge that modifies the gravitational potential generated
by the atmosphere.
The resulting tidal torque is
\citep{Dobrovolskis_Ingersoll_1980, Correia_Laskar_2003JGR}:
\begin{eqnarray}
T_a & = & \frac{3}{2} K_a \left[ \left(1 \aa e^2\right) \, b_a (2
\omega - 2 n)  \right. \crm
& & \left. \ab \, e^2 b_a (2 \omega - n) \ac \, e^2 b_a (2 \omega - 3 n)
\right] \ , \label{eq02b}
\end{eqnarray}
where $ K_a = - 3 M_* R^3 / (5 \rho a^3) $,
and $ \rho $ is the mean density of the planet.
The amplitude of the bulge is given by $ \tilde p_2 $, the second order surface
pressure variations \citep{Chapman_Lindzen_1970}:
\begin{equation}
\tilde{p}_2 (\sigma) = \mathrm{i} \frac{\gamma}{\sigma} \tilde p_0 \left(
\nabla \cdot \vec{v}_\sigma - \frac{\gamma - 1}{\gamma} \frac{J_\sigma}{g H_0}
\right) = \mathrm{i} \frac{\cal P_\sigma}{\sigma} \ , \label{eq06}
\end{equation}
where $ \gamma = 7/5 $ for a perfect diatomic gas, $ \tilde p_0 $ is the mean surface
pressure, $ \vec{v} $ is the velocity of tidal
winds, $ J $ is the amount of heat absorbed or emitted by a unit mass of air per
unit time, and $ H_0 $ is the scale height at the surface.
There is also a delay $ \Delta t_a $ before the response of the atmosphere to
the stellar heat excitation.
However, the imaginary number in Eq.(\ref{eq06}) causes the pressure variations to
lead the Sun ($ \mathrm{i} = e^{\mathrm{i} \pi / 2} $).
Using $ \delta_a = \sigma \Delta t_a /2 $, we thus have
$ 
b_a (\sigma) = \tilde p_2 \sin 2 \delta_a  = | \tilde p_2 | \sin 2 ( \delta_a +
\pi / 2) = - | \tilde p_2 | \sin 2 \delta_a 
$. 




The dependence of time lags on the tidal frequency is poorly known
for gravitational and atmospheric tides.
For a review of the different dissipation models, see
\citet{Correia_etal_2003}.
For both tides we adopt 
the simplest model for slow rotation (i.e. the viscous model), as
described by \citet{Mignard_1979}, where
a constant time lag is assumed  for all components of the tidal perturbation.
Since we usually have $ \sigma \Delta t \ll 1 $, this model can be made linear:~
$ 
b_g(\sigma) \simeq k_2 \, \sigma \Delta t_g \quad \mathrm{and} \quad
b_a(\sigma) \simeq - | \tilde p_2 | \, \sigma \Delta t_a 
$. 
For atmospheric tides, it is also necessary to consider the response of the
surface pressure variations to tidal frequency (Eq.\,\ref{eq06}).
We use the ``heating-at-the-ground model'' described by \citet{Dobrovolskis_Ingersoll_1980}.
It is supposed that all the stellar flux absorbed by the ground, $ F_s $, is
immediately deposited in a thin layer of atmosphere at the surface. 
The heating distributing is then written as a delta-function just above
the ground ($ J_\sigma = g F_s / \tilde p_0 $).
This approximation is justified because tides in the upper atmosphere are
decoupled from the ground by the disparity between their rotation rates and it
appears to be in good agreement with the observations.
Neglecting \vec{v_\sigma} over the thin heated layer, Eq.(\ref{eq06})
becomes:
$ 
{\cal P_\sigma} = F_s / (8 H_0) \propto L_* / a^2 
$ 
($ L_* $ is the star luminosity).


The average evolution of the rotation rate is obtained by adding the effects
of both tidal
torques acting on the planet, that is $ \dot \omega = (T_a + T_g) / C $.
Substituting dissipative models described above in Eqs.(\ref{eq02}) and
(\ref{eq02b}) then leads to 
\begin{eqnarray}
\dot \omega / K_0 & = &  \omega - \left(1 + 6 e^2\right) n - \omega_s
\left[ \left(1 \aa e^2\right) \sign (\omega - n)  \right.  \crm
& & \left. \ab \, e^2 \sign (2 \omega - n) \ac \, e^2 \sign (2 \omega - 3 n)
\right] \ , \label{eq53}
\end{eqnarray}
where $ K_0 = - 3 \Omega(e) K_g k_2 \Delta t_g / C $, $ \; \, \Omega (e) = 1 +
15 e^2/2 $,
\begin{equation}
\omega_s = \frac{F_s}{8  H_0 \Omega(e) k_2} \frac{K_a \Delta t_a}{K_g \Delta
t_g} \propto \frac{L_*}{M_*} \, \frac{R}{m} \, a \ , \label{eq54}
\end{equation}
and $ m $ is the planet mass.
According to Eq.(\ref{eq06}), atmospheric tides are weak during the first stages
of evolution ($ \sigma \gg n $).
Gravitational tides alone can therefore be used to estimate the characteristic
time needed to reach the equilibrium rotation, $ \tau_{eq} $:
\begin{equation}
\tau_{eq}^{-1} \sim \left| K_0 \right| = \frac{3
k_2\, \Delta t_g\, R^5}{C\,G } \, n^4 \sim  \frac{9\,
G \,M_*^2\, k_2\, \Delta t_g\, R^3}{m\,a^6 } \ , \label{eq21}
\end{equation}
with $ C \simeq m\, R^2 / 3 $.
All terrestrial extra-solar planets
listed in Table\,\ref{Tab1} have a despinning timescale that is significantly
lower than the age of the system ($< 10^7$~yrs), so we
expect that they have already reached their equilibrium rotation state.

\section{Equilibrium final states for the rotation rate}

\begin{figure}
\begin{center}
\begin{tabular}{c}
\includegraphics[width=7.9cm]{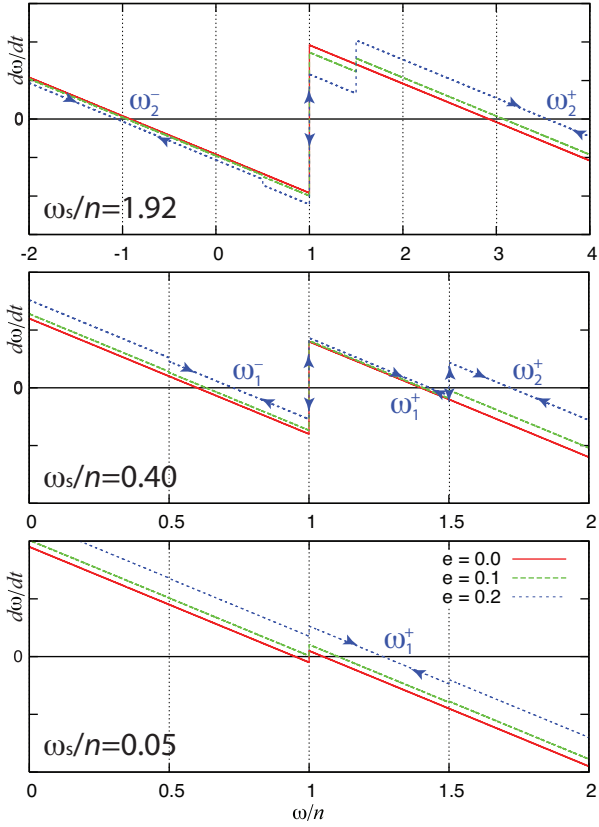} 
\end{tabular}
\caption{Evolution of $ \dot \omega $ (Eq.\,\ref{eq53}) with $ \omega_s /
n = 1.92 $, $ \omega_s / n = 0.40 $ and $ \omega_s / n = 0.05 $,  for different
eccentricities ($ e = 0.0, 0.1, 0.2 $).
The equilibrium rotation rates are given by $ \dot \omega = 0 $ and the arrows
indicate whether it is a stable or unstable equilibrium position.
For $ \omega_s / n > 1 $, we have two equilibrium possibilities, $
\omega^\pm_2 $, one of which corresponds to a retrograde rotation (as
for Venus).
For $ \omega_s / n < 1 $, retrograde states are not possible, but we can still
observe final rotation rates $ \omega^- < n $.
For eccentric orbits, because of the terms in $ b_\tau (2 \omega - n) $ and $
b_\tau (2 \omega - 3 n) $, we may have at most four different final
possibilities (Eq.\ref{eq58a}).
When $ \omega_s /n $ becomes~extremely small, which is the case for the present
observed extra-solar planets with some eccentricity (Table\,\ref{Tab1}), a single
final equilibrium is possible for $ \omega^+_1 $. \label{Fig1}}
\end{center}
\end{figure}

An equilibrium final state is achieved when $ \dot \omega = 0 $, that
is for $ T_g = - T_a $.
When $ e = 0 $, we derive:
\begin{equation}
f(\omega - n) = - \frac{T_g}{T_a} = - \frac{K_g b_g (2 \omega - 2 n)}{K_a b_a (2 \omega - 2 n)} = 1
\ , \label{eq14}
\end{equation}
where $ f(x) $ is an even function of $ x $ \citep{Correia_Laskar_2001}.
Assuming that $ f (x) $ is monotonic close to the equilibrium 
(which is true for the usual dissipation models), we have 
\begin{equation}
| \omega - n | = f^{-1} (1) \equiv \omega_s \ , \label{eq15}
\end{equation}
i.e. there are two final possibilities for the equilibrium rotation of the
planet, given by $ \omega^\pm = n \pm \omega_s $, where $ 2 \pi / \omega_s $ can
be seen as the synodic period.
Unless $ \omega_s \ll n $, the synchronous motion ($ \omega = n $) is no
longer the final equilibrium position for Earth-like planets tidally evolved.
According to \citet{Correia_Laskar_2001, Correia_Laskar_2003I}, the planet is
free to evolve to any of the two final equilibrium positions, although it
appears to have a preference for the slowest one ($ \omega^- = n - \omega_s $).
If $ \omega_s < n $, both final states correspond to prograde final rotation
rates (Fig.\ref{Fig1}b,c).
However, if $ \omega_s > n $, the slowest rotation rate becomes negative,
corresponding to a retrograde configuration. 
This is the present situation of
the planet Venus for which $ \omega_s /n = 1.92 $ (Fig.\ref{Fig1}a).

When $ e \ne 0 $, Eq.(\ref{eq15}) is no longer valid and additional equilibrium
positions for the rotation rate may occur.
For moderate values of the eccentricity, we can write from Eqs.(\ref{eq02}) and
(\ref{eq02b}):
\begin{equation}
f(\omega - n) = 1 + e^2 \left(\ar + \frac{g(\omega) }{b_a (2 \omega - 2 n)} \right)
 \ , \label{eq56}
\end{equation}
where
\begin{eqnarray}
g(\omega) & = & \frac{K_g}{4 K_a} \left[ b_g(2 \omega - n) + 49 b_g (2 \omega -
3 n) \right]  \crm & & \ab \, b_a (2 \omega - n) \ac \, b_a (2 \omega - 3 n) \ .
\label{eq57}
\end{eqnarray}
Since $ b_\tau (\sigma) $ are monotonic odd functions, the effect of the
eccentricity is eventually to split each previous equilibrium rotation rate into
two new equilibrium values so that four final equilibrium positions for the
rotation rate are possible, written as:
\begin{equation}
\omega^\pm_{1,2} = n \pm \omega_s + e^2 \, \dw^\pm_{1,2} \ , \label{eq58a}
\end{equation}
with
\begin{equation}
\dw^\pm_{1,2} = \left( \ar + \frac{g(\omega)}{| b_a(2 \omega - 2 n)|}
\right) \left. \frac{\partial f^{-1}}{\partial x} \right|_{x=1} \ ,
\label{eq58b}
\end{equation}
or, adopting the tidal models described in Sect.~\ref{tmodels}
(Eq.\,\ref{eq53}):
\begin{equation}
\dw^-_{1,2} =  6 n - \left( 6 \pm 1 \right) \omega_s \quad \mathrm{and} \quad
\dw^+_{1,2} =  6 n - \left( 4 \pm 9 \right) \omega_s \ , \label{eq59}
\end{equation}
where $ + $ corresponds to the state $ \delta_1 $ and $-$ to the state $
\delta_2 $. Because the set of $\omega^\pm_{1,2}$ values must verify the additional condition
\begin{equation}
\omega^-_2 < n/2 < \omega^-_1 < n < \omega^+_1 < 3 n / 2 <  \omega^+_2 \ ,
\label{cond}
\end{equation}
these four equilibrium rotation states cannot, in general, exist simultaneously,
depending on the values of $\omega_s$ and $ e $.
In particular, the final states $\omega^-_1$ and $\omega^+_1$
can never coexist with $ \omega^-_2 $. 
At most three different equilibrium states are therefore possible, obtained when
$\omega_s/n$ is close to $1/2$, or more precisely, when $1/2-17\,e^2/2 < \omega_s / n < 1/2 + 7\,e^2/2$.
Conversely, we found that one single final state $\omega^+_1 = (1 + 6 e^2 ) \, n
+ (1 - 13 e^2) \, \omega_s $ exists when $\omega_s / n < 6 e^2 ( 1 - 7 e^2 )$.


\begin{figure*}
\begin{center}
\begin{tabular}{c}
\includegraphics[width=16.8cm]{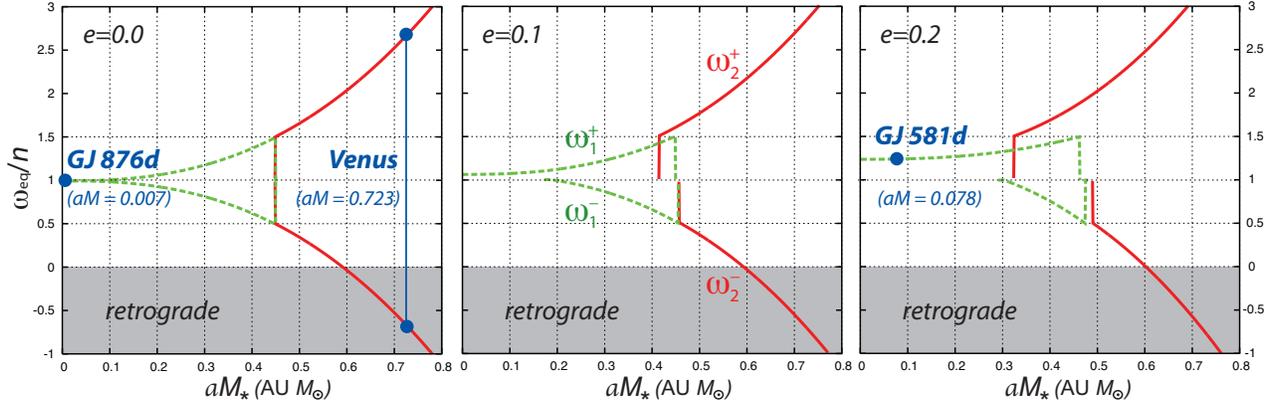} 
\end{tabular}
\caption{\bfi Equilibrium positions of the rotation rate as a function of the product
$ a M_* $ for three different values of the eccentricity ($e = 0.0, 0.1, 0.2 $).
Each curve corresponds to a different final state (dotted lines for $\omega_1^\pm
$ and solid lines for $\omega_2^\pm $).
For $ e \simeq 0 $ (case of Venus), we always count two final states that are symmetrical
about $ n $. For small values of $ a M_* $, the two equilibrium
possibilities are so close to $ n $ that the most likely scenario for the planet is to
be captured in the synchronous resonance (case of {\small GJ}\,876\,d).
As we increase the eccentricity, we can count at most three final equilibrium rotations,
depending on the value of $\omega_s/n$ (computed from Eq.\,\ref{eq60}).
When $ e \simeq 0.2 $, only one equilibrium state exists for $ a M_* < 0.3 $,
resulting from $ \omega_s / n < 6 e^2 ( 1 - 7 e^2 ) $. This is 
the present situation of {\small GJ}\,581\,d and most of the ``Earth-like''
extra-solar planets listed in Table\,\ref{Tab1}.} 
\label{Fig2}
\end{center}
\end{figure*}



\begin{table*}
\centering
\caption[]{Characteristics and equilibrium rotation rates of Earth-like planets
with masses lower than
$12 \, M_{\oplus}$ (see text for notations).}
\label{Tab1}
\begin{tabular}{lcccccc|cccccc}
\hline
\noalign{\smallskip}
 Name & $M_*$ & Age & $^* \tau_{eq}$ & $ m \sin i $ & $a$ & $ e $ &
 $ \omega_s / n $ & $ 2 \pi /n $ & $ 2 \pi / \omega^-_1 $ &
 $ 2 \pi / \omega^-_2 $ & $ 2 \pi / \omega^+_1 $ & $ 2 \pi / \omega^+_2 $ \\

 & [$M_\odot$] & [Gyr] & [Gyr] & $[m_\oplus]$ & [AU]  & & &
 [day] & [day]  & [day]  & [day] & [day] \\

\hline
\noalign{\smallskip}
Venus & 1.00 & 4.5 & 2.3 & 0.82 & 0.723 & 0.007 &
1.92 & 224.7 &   & $-$243 &   & 76.8 \\


GJ\,581\,c$^{1}$ & 0.31 & 4.3 & $10^{-5}$ & 5.0 & 0.073 & 0.16 &
0.0002 & 12.93 &   &   & 11.2 &   \\

GJ\,876\,d$^{2}$ & 0.32 & 9.9 & $10^{-8}$ & 5.7 & 0.021 & 0 &
$10^{-5} $ & 1.9378 & 1.9379 &   & 1.9377 &   \\

GJ\,581\,d$^{1}$ & 0.31 & 4.3 & 0.04 & 7.7 & 0.253 & 0.2 &
0.0026 & 83.6 &   &   & 67.4 &   \\

HD\,69830\,b$^{3}$ & 0.86 & 4-10 & $10^{-5}$ & 10.2 & 0.079 & 0.10 &
0.0009 & 8.667 &   &   & 8.17 &   \\

GJ\,674\,b$^{4}$ & 0.35 & 0.1-1 & $10^{-7}$ & 11.7 & 0.039 & 0.2 &
$10^{-5} $ & 4.693 &   &   & 3.79 &   \\

HD\,69830\,c$^{3}$ & 0.86 & 4-10 & $10^{-3} $ & 11.8 & 0.186 & 0.13 &
0.0069 & 31.56 &   &   & 28.5 &   \\

 \noalign{\smallskip}
 \hline
 \noalign{\smallskip}
 \end{tabular}

$^*$ Using Eq.(\ref{eq21}) with $ k_2 = 1/3 $ and $ \Delta t_g = 640 $\,s (Earth's
values);
References: [1] \citet{Udry_etal_2007}; [2]
\citet{Rivera_etal_2005}; [3] \citet{Lovis_etal_2006}; [4] \citet{Bonfils_etal_2007}.
Earth-like planets OGLE-2005-BLG-390Lb \citep{Beaulieu_etal_2006} and
MOA-2007-BLG-192-Lb \citep{Bennett_etal_2008} have not been included because 
 their despinning timescales $\tau_{eq} \sim 10^2 $\,Gyr are much
 larger than the age of the Universe. 

\end{table*}

\section{Application to Earth-like extra-solar planets}

The Earth and Venus are the only Earth-like planets for which the atmosphere and spin
are known.
Only Venus is tidally evolved and therefore suitable for applying the above
expressions for tidal equilibrium.
We can nevertheless investigate the final equilibrium rotation states of
the already detected ``super-Earths''.
For that purpose, we considered only the 6 extra-solar planets of masses smaller than 12\,$M_{\oplus}$ that we
 classified as rocky planets with a dense atmosphere, although we stress that this mass boundary is quite arbitrarily.
Using the empirical mass-luminosity relation $ L_* \propto M_*^{4} $
\citep[eg.][]{Cester_etal_1983} and the mass-radius
relationship for terrestrial planets $ R \propto m^{0.274} $ \citep{Sotin_etal_2007},
 Eq.(\ref{eq54}) can be written as:

\begin{equation}
\omega_s / n = k \, (a \, M_*)^{2.5}  m^{-0.726} \ , \label{eq60}
\end{equation}
where $ k $ is a proportionality coefficient that contains all the constant
parameters, but also the parameters that we are unable to
constrain such as $ H_0 $, $ k_2 $, $ \Delta t_g $ or $ \Delta t_a $.
In this context, as a first order approximation, 
we consider that for all these terrestrial planets,  the parameter $k$ 
has the same value as for Venus.
Assuming that the rotation of Venus is presently stabilized in the $ \omega^- $
final state, that is, $ 2 \pi / \omega^- = - 243 $\,days
\citep{Carpenter_1970}, we compute $ 2 \pi / \omega_s = 116.7 $\,days.
Replacing it in Eq.(\ref{eq60}), we find for Venus that
$ k = 3.32 \; m_\oplus^{0.726} M_\odot^{-2.5} $AU$^{-2.5}$.
We can then estimate the ratio $ \omega_s/n $ for all considered extra-solar
planets in order to derive their respective equilibrium rotation rates (Table\,\ref{Tab1}).
The number and values of the equilibrium rotation states are plotted as a function
of $a M_* $ for different eccentricities in Fig.\,\ref{Fig2}.
All eccentric planets have a ratio $\omega_s/n$ that is lower than $7 \times 10^{-3}$,
which verifies the condition  $ \omega_s / n < 6 e^2 (1-7 e^2 ) $. As a consequence,
only one single final state $\omega^+_1 / n \simeq (1+6\,e^2) $ exists, corresponding to the equilibrium
rotation induced by gravitational tides only (Eq.\,\ref{eq02}). The main reason
is that the effect of atmospheric tides is clearly disfavored relative to the
effect of gravitational tides on Earth-like planets discovered orbiting
M-dwarf stars:
The smaller orbital distance strengthens the effect of gravitational tides, which
are proportional to $1/a^6$, while the effect of thermal tides varies as $1/a^5$. 
Moreover, the smaller mass of the
central star also strongly affects the luminosity received by the planet and hence
the size of the atmospheric bulge driven by thermal contrasts.

The only exception is {\small GJ}\,876\,d, because its eccentricity is believed
to be zero (yet to be confirmed). However, the two equilibrium rotation states
$\omega^\pm_1 $ are so
close to the mean motion $ n $, that the quadrupole moment of inertia (not
included in our analysis) will probably capture the rotation in
synchronous resonance.

\section{Discussion and conclusion}

We have derived a simple model that permits the determination of the final
equilibrium rotation of ``Earth-like'' planets.
Our model contains some uncertain parameters related to the dissipation within
the planets, but we were able to gather all this information in a single
parameter, $\omega_s$.
By varying this, we can then cover all possibilities for the
rotation of terrestrial planets. We demonstrated that for a planet of moderate
eccentricity and low obliquity, at most four final
equilibrium positions are possible. 
For eccentricities higher than $ e \sim 0.5 $,
terms of higher degree in $e$ should be considered in Eq.(\ref{eq02}) 
that may generate additional equilibrium positions.

Based on the present rotation of Venus, we provided an estimate of $\omega_s$
for different environments (Eq.\,\ref{eq60}).
An important consequence is that the ratio $ \omega_s / n $ increases rapidly
with the semi-major axis and mass of the star because $ \omega_s /n
\propto (a M_*)^{2.5} $. The effect of the atmosphere on the equilibrium
rotation is therefore more
relevant for planets that orbit Sun-like stars at not close distances.
Unfortunately, this situation is precisely the opposite of that for the
discovered Earth-like planets (Table\,\ref{Tab1}), since the radial velocity technique is more sensitive to
the detection of short-period planets.
{\bfi As a consequence, unlike Venus, none of these planets can be stabilized with a
rotation rate $\omega < n $, because for $ e > 0.1 $ these final states only
exist if $ a M_* > 0.2 $\,AU\,$M_\odot$ (Fig.\ref{Fig2}).}
The effect of atmospheric tides is extremely small ($ \omega_s / n \sim 0 $)
and the equilibrium rotation is essentially driven by the
tidal gravitational torque.
However, their significant eccentricity ($0.1 < e < 0.2$) moves the equilibrium position
away from the synchronous motion.
Capture in this resonance is unlikely, but capture in a higher order
spin-orbit resonance cannot be rejected, in particular for {\small GJ}\,581\,d
and {\small GJ}\,674\,b, depending on the asymmetry of their mass distribution.


\begin{acknowledgements}
We thank F. Selsis for useful discussions. This work was supported by the
Funda\c{c}\~{a}o Calouste Gulbenkian (Portugal), 
Funda\c{c}\~{a}o para a Ci\^{e}ncia e a Tecnologia (Portugal), and by PNP-CNRS
(France).
\end{acknowledgements}

\bibliographystyle{aa}
\bibliography{correia}

\end{document}